\documentclass[twocolumn,prb]{revtex4}
\usepackage{epsfig,amssymb,amsmath}
\begin{document}
\title{Effect of coupling between junctions on superconducting current and charge correlations in intrinsic Josephson junctions}
\author{Yu. M. Shukrinov,$^{1}$ M. Hamdipour,$^{1,2}$ and M. R. Kolahchi$^{2}$}

\affiliation{$^{1}$ BLTP, JINR, Dubna, Moscow Region, 141980, Russia}

\affiliation{$^{2}$Institute for Advanced Studies in Basic Sciences, P.O. Box 45195-1159, Zanjan, Iran}
\date{\today}
\begin{abstract}
Charge formations on superconducting layers and creation of the longitudinal plasma wave in the stack of
intrinsic Josephson junctions change crucially the superconducting current through the stack. Investigation of
the correlations of superconducting currents in neighboring Josephson junctions and the charge correlations in
neighboring superconducting layers allows us to predict the additional features in the current--voltage
characteristics. The charge autocorrelation functions clearly demonstrate the difference between harmonic and
chaotic behavior in the breakpoint region.  Use of the correlation functions gives us a powerful method for the
analysis of the current--voltage characteristics of coupled Josephson junctions.
\end{abstract}
\maketitle The intrinsic Josephson junctions (IJJ) have a wide interest today due to the observed powerful
coherent radiation from the stack of IJJ.\cite{ozyuzer} The radiation is related to the region in the
current--voltage characteristics (CVC) closed to the breakpoint region (BPR).\cite{sust1,prl,prb} The
resistively and capacitively shunted junction (RCSJ) model and its different modifications are well known to
describe the properties of single Josephson junctions, giving a clear picture of the role of quasiparticle and
superconducting currents in the formation of CVC.\cite{kleiner,likharev,barone} In the case of a stack of IJJ
the situation is cardinally different. The system of the coupled Josephson junctions has a multiple branch
structure and it has additional characteristics: the breakpoint current, the transition current to another
branch and the BPR width. The breakpoint features were predicted theoretically and observed experimentally
recently.\cite{irie} We demonstrated that the CVC of the stack exhibits a fine structure in the BPR.\cite{prb2}
The breakpoint manifests itself in the numerical simulations of the other authors as well.\cite{machida99}

The breakpoint current characterizes  the resonance point, at which the longitudinal plasma wave (LPW) is
created in stacks, with a given number and distribution of the rotating and oscillating IJJ.  These notions
should be taken into account to have a correct interpretation of the experimental results.  The investigation of
the coupled system of Josephson junctions with  a small value of the coupling parameter (as in the case of
capacitive coupling), allowed us to understand in a significant way, the influence of the coupling between
junctions on physical properties of the system. The capacitive coupling is realized in nanojunctions if the
length of the junction is comparable to, or smaller than the Josephson penetration depth at zero external
magnetic field. Coupling between intrinsic Josephson junctions leads to the interesting features which are
absent in single Josephson junction. Still, the superconducting current in the coupled system of Josephson
junctions with LPW has not been investigated in detail; this includes the role of the correlations of the
superconducting currents in different junctions and the charge correlations on superconducting layers.

Here we study the phase dynamics of an IJJ stack in high-$T_c$ superconductor. The CVC of IJJ are numerically
calculated in the framework of capacitively coupled Josephson junctions model with diffusion
current.\cite{machida00,physC2} We find that the behavior of the superconducting current in the coupled system
of Josephson junctions is essentially different from that of a single Josephson junction. It is demonstrated
that superconducting current in the stack of IJJ reflects the main features of the breakpoint region; in
particular, the fine structure in the CVC. We study various correlation functions in the characteristics, and
observe that the correlations amongst the charge on different superconducting layers and correlations in the
superconducting currents of different junctions lead to the detailed features of CVC in the BPR, and also
provides additional information about the phase dynamics in the IJJ.

\begin{figure}[ht]
 \centering
\includegraphics[height=75mm]{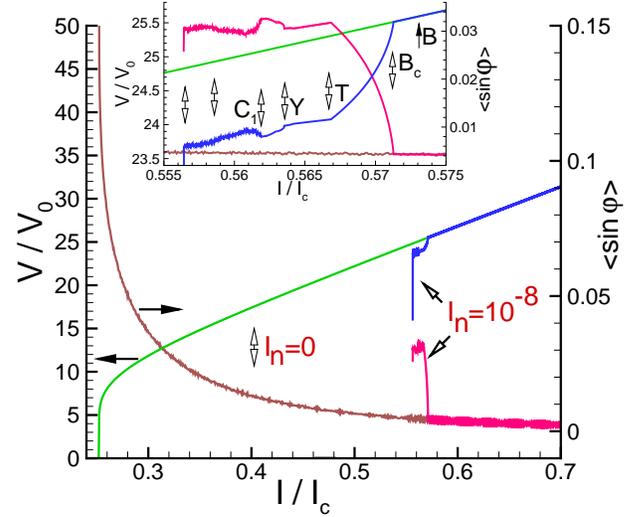}
\caption{(Color online) The CVC and time averaged superconducting current $<\sin \phi_l>$ with
 no noise in
current ($I_n=0$), and with noise amplitude $I_n=10^{-8}$. The inset shows an enlarged view of the BPR. }
 \label{1}
\end{figure}

To find the CVC of the stack with N IJJ, we solve a system of N dynamical equations for the gauge--invariant
phase differences $\varphi_l(t)\equiv \varphi_{l,l+1}(t)= \theta_{l+1}(t)-\theta_{l}(t)-\frac{2e}{\hbar
c}\int^{l+1}_{l}dz A_{z}(z,t)$ between superconducting layers ($S$-layers),  where $\theta_{l}$ is the phase of
the order parameter in the S-layer $l$,  and $A_z$ is the vector potential in the barrier. In our simulations we
use a dimensionless time $\tau = t\omega_p$, where $\omega_{p}$ is the plasma frequency
$\omega_{p}=\sqrt{2eI_c/\hbar C}$, ${I_c}$ is the critical current, and $C$ is the capacitance. The voltage is
presented in units of $V_0=\hbar\omega_p/(2e)$, and the current in units of $I_c$. The system of equations has a
form $\partial^2\varphi_l/\partial \tau^2 = \sum\sb{l'}\,A_{ll'}[I-\sin\varphi_{l'}-\beta\partial
\varphi_{l'}/\partial \tau]$ with matrix $A$ given in Ref. 4, for periodic and nonperiodic boundary conditions
(BC). Using Maxwell's equation $\emph{div} (\varepsilon E) = \rho/\varepsilon_0$, where $\varepsilon$ and
$\varepsilon_0$ are relative dielectric and electric constants, we express the charge density $Q_l$ (we call it
just charge) in the S-layer $l$ by the voltages $V_{l}$ and $V_{l+1}$ in the neighboring insulating layers
$Q_l=Q_0 \alpha (V_{l+1}-V_{l})$, where $Q_0 = \varepsilon \varepsilon _0 V_0/r_D^2$,  and $r_D$ is the Debye
screening length. Solution of the system of dynamical equations for the gauge--invariant phase differences
between S-layers gives us the voltages $V_{l}$ in all junctions in the stack, and allows us to investigate the
time dependence of the charge on each S-layer. The time dependence of the charge consists of time and bias
current variations. We solve the system of dynamical equations for phase differences at fixed value of bias
current $\emph{I}$ in some time $\tau$ domain $(0, T_m)$ with a time step $\delta \tau$; we then change the bias
current by the current step $\delta I$, and repeat the same procedure for the current $I+\delta I$ in the new
time interval $(T_m, 2T_m)$. The values of the phase and its time derivative at the end of the first time
interval, are used as the initial conditions for the second time interval and so on. In our simulations we use
$T_m=25000$, $\delta \tau=0.05$, $\delta I=10^{-5}$ and the dimensionless time $t_r$ is recorded as
$t_r=\tau+T_m(I_0 -I)/\delta I$, where $I_0$ is an initial value of the bias current. In this paper the CVC and
the time dependence of charge oscillations in the S-layers are simulated at  $\alpha = 1$, $\beta = 0.2$, using
periodic BC. The details concerning the model and the numerical procedures have been presented before.
\cite{physC2,prl,prb}

\begin{figure}[ht]
 \centering
\includegraphics[height=75mm]{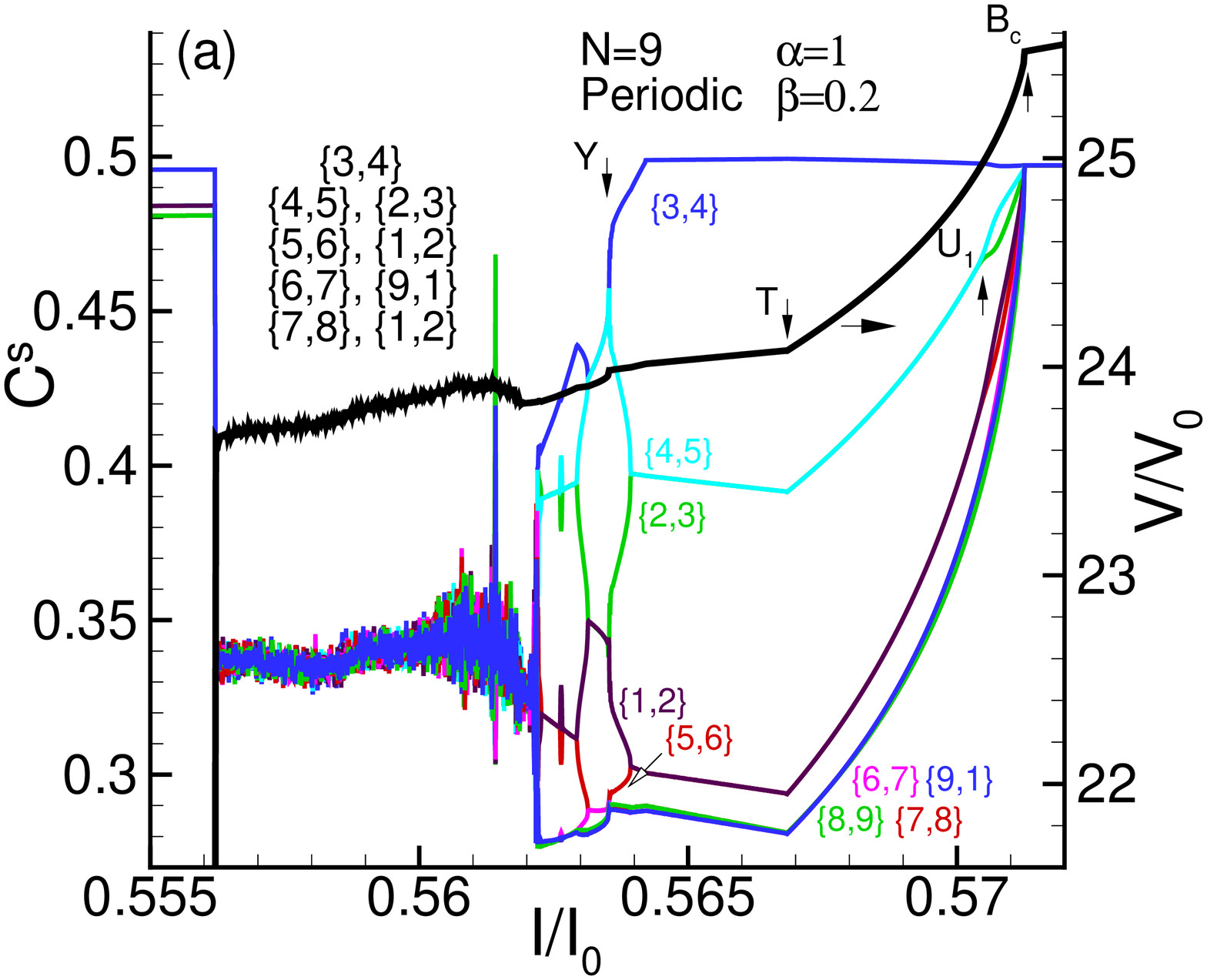}
\includegraphics[height=75mm]{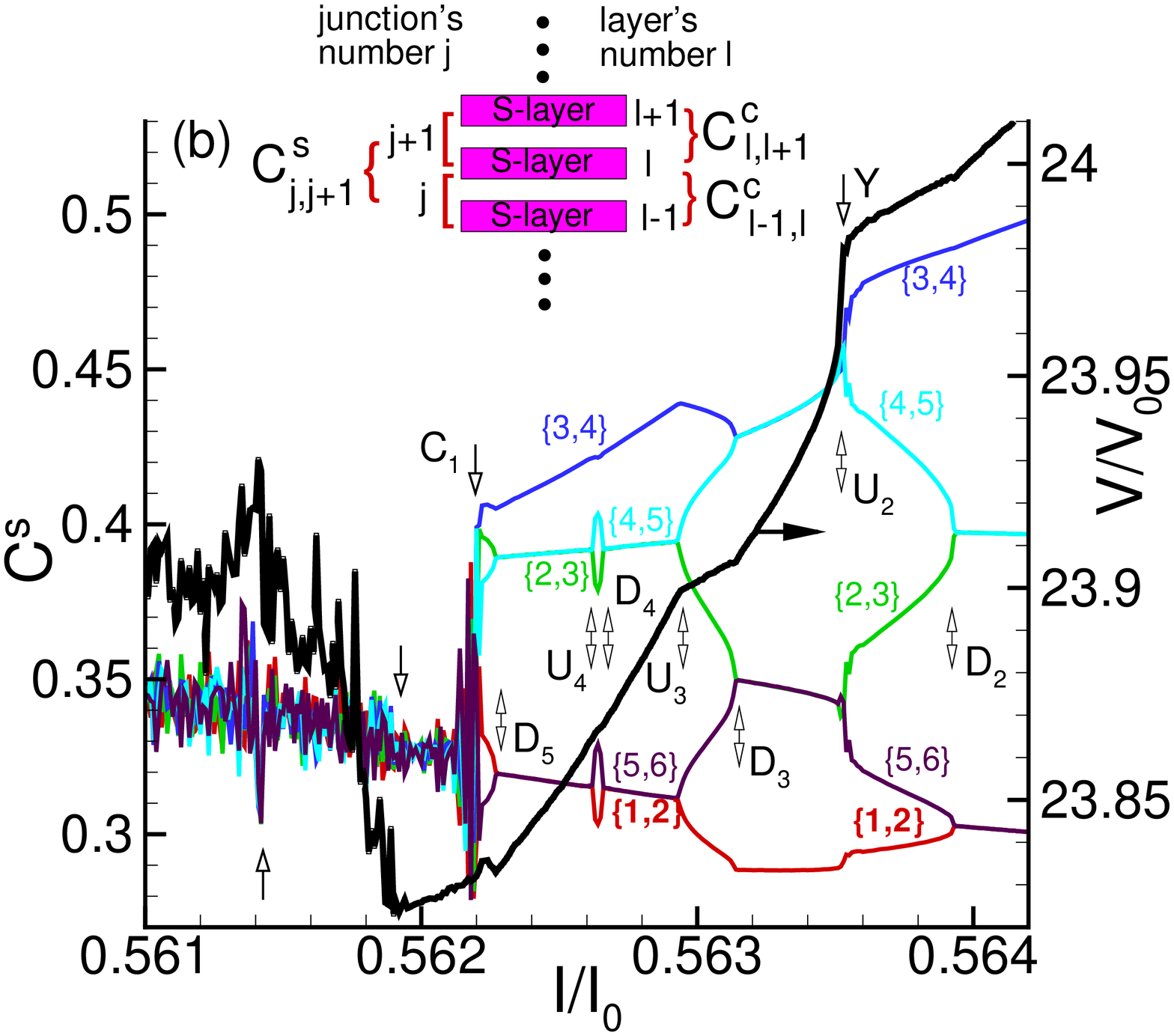}
\caption{(Color online) (a) The correlation of superconducting currents in the stack with nine IJJ: the color
curves plot the correlation functions $C^{s}_{j,j+1} = <\sin \varphi_j(\tau)\sin \varphi_{j+1}(\tau)>$  as a
function of bias current for all $j$, the black curve shows the corresponding CVC. Label $\{2,3\}$, for
instance, indicates the charge or current correlations between layers $2$ and $3$. (b) The enlarged region in
the current interval (0.561, 0.564). Notations for current and charge correlation functions are also given in
the inset.}
 \label{2}
\end{figure}

Fig.~\ref{1} shows the CVC (green and blue curves) and the time averaged superconducting current $<\sin \phi_l>$
(brown and red curves ) without random noise in current,  and with noise amplitude $10^{-8}$ for stack with nine
IJJ. Our simulations of the CVC without noise gives us the value of the return current as $I_r/I_c=0.2517$,
which coinciding with the value obtained from the RCSJ model. In fact, in this model the relation between the
return and critical currents for $\beta<<1$ has the form $I_r/I_c=4\beta/ \pi$, so that at $\beta=0.2$ we get
$I_r/I_c=0.2546$. The superconducting current without noise  demonstrates the standard increase before
transition to the zero voltage state.\cite{likharev} The noise in current helps create the LPW in the stack and
influences the superconducting current; the particular value of noise is not very important. The creation of the
LPW in the stack of junctions changes the CVC drastically, leading to the breakpoint and breakpoint region in
CVC. Inset shows the enlarged part of the Fig.~\ref{1} in the breakpoint region, where arrows indicate the
coincidence of the main features of CVC and the superconducting current. The point $B$ shows the point where a
charge appears on the S-layers; point $B_c$ is a breakpoint on the CVC and it reflects the breakdown in the
sharp increase of the charge's value.\cite{prb2} If we take a sum of all equations
$\ddot{\varphi_l}=(1-\alpha\triangledown^{2}_{l})(I-\beta\dot{\varphi_l}-\sin\varphi_l)$ for N IJJ and then find
its average in time, we obtain the equation $V=\frac{N}{\beta}(I-<sin\varphi>)$. This equation clearly shows why
our results; namely, the $V$-curve and the $<sin\varphi>$-curve,  demonstrate the same features.

To investigate the origin of the CVC features in the BPR, we study the correlations, $C^{s}_{j,j+1}$, of
superconducting currents in the neighboring junctions $j$ and $j+1$: $C^{s}_{j,j+1} = <\sin \varphi_j(\tau)\sin
\varphi_{j+1}(\tau)>= \lim_{(T_m-T_i)\rightarrow\infty} \frac{1}{(T_m-T_i)} \int_{T_i}^{T_m}{\sin
\varphi_j(\tau)\sin \varphi_{j+1}(\tau)d \tau}$, where the brackets $<>$ mean averaging over time. The
$C^{s}_{j,j+1}$  as functions of bias current $I/I_c$ are presented in Fig.~\ref{2}a for $j=1,...,9$. All curves
practically coincide in the interval starting from the breakpoint $B$ till point $B_c$. Then we observe regular,
but different, behavior for different $j$ till point $C_1$, and then again $C^{s}_{j,j+1}$ is nearly the same
for all $j$ in the chaotic region. As we can see in Fig.~\ref{2}a,  eight of these functions come in pairs as
$(C^{s}_{4,5},C^{s}_{2,3})$, $(C^{s}_{5,6},C^{s}_{1,2})$, $(C^{s}_{6,7},C^{s}_{9,1})$,
$(C^{s}_{7,8},C^{s}_{1,2})$; and one function, $C^{s}_{3,4}$ stands by itself.  Let us note that because we
consider a periodic BC, the number assigned to a junction is only a label.  The black curve shows the outermost
branch of the CVC of the stack with nine IJJ. We can see that the features of the correlation functions coincide
with the features of CVC, i.e. they manifest themselves in the CVC curve. In Fig.~\ref{2}b the part of
Fig.~\ref{2}a is plotted on an expanded scale. Arrows show the points where the features of $C^{s}_{j,j+1}$
coincide with those of CVC. In the inset to this figure we clarify the formation of the charge and
superconducting current correlation functions by the diagram. Study of $C^{s}_{j,j+1}$ allows us to find new
features in CVC which were not noticed in the previous studies.\cite{prb2} Particularly, in Fig.~\ref{2} we
indicate the points $D_2$, $D_3$, $D_4$, and $D_5$ where the curves of the correlation functions diverge, and
points $U_1$, $U_2$, $U_3$, $U_4$, where they converge.

\begin{figure}[ht]
 \centering
\includegraphics[height=70mm]{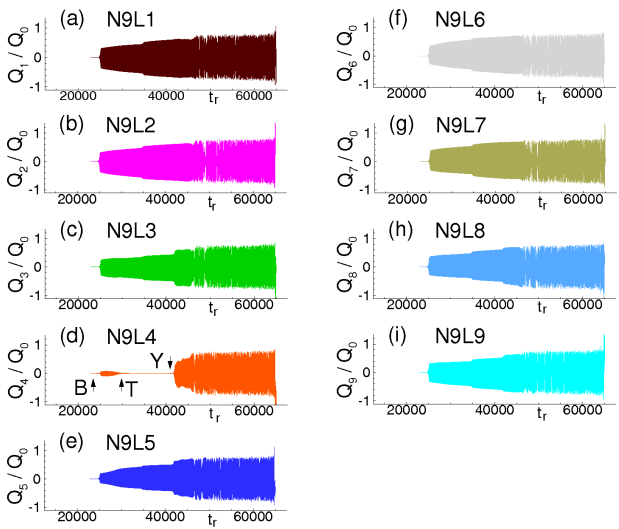}
\caption{(Color online)Profile of time dependence of the charge in the S-layers in the stack with 9 IJJ at
$\alpha=1$, $\beta=0.2$ and with periodic BC.}
 \label{3}
\end{figure}

\begin{figure}[ht]
 \centering
\includegraphics[height=75mm]{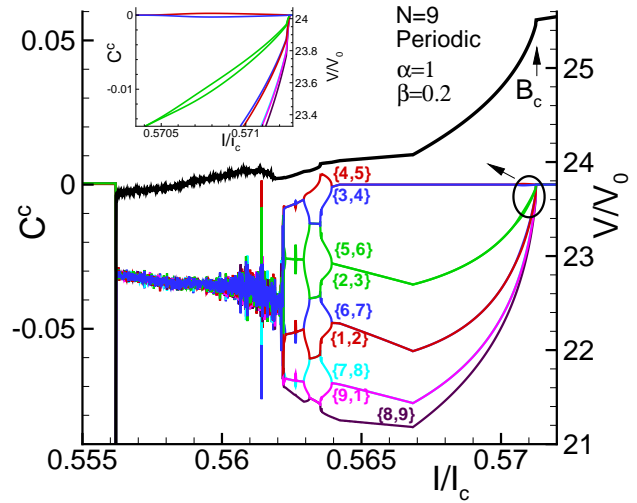}
\caption{(Color online)  The correlation of charge in the stack with nine IJJ: the color curves plot the
correlation functions $C^{q}_{l,l+1}=<Q_l(\tau) Q_{l+1}(\tau))>$ for all $l$ as a function of bias current; the
black curve shows the corresponding CVC.  The inset shows the enlarged region near the point $B_c$.}
 \label{4}
\end{figure}

To understand the origin of these features, we study the time dependence of charge in the superconducting
layers.   In Fig.~\ref{3} we show the profiles of the charge oscillations in all layers for the stack of nine
IJJ. The odd number of junctions in the stack at periodic BC leads to the case where the charge dynamics in one
layer, qualitatively differs from the others; namely, the layer 4 demonstrates a specific time dependence of the
charge oscillations. We will refer to this layer as the  ``specific layer" or the sp-layer for short. The value
of the charge on the sp-layer is  smaller than on the other layers up to point $Y$, and  practically staying at
zero, in the interval from point $T$ till point $Y$.  Such difference leads to the maximal value of
$C^{s}_{3,4}$. The correlation function $C^{s}_{3,4}$ is different from the other correlation functions, because
both phase differences in it ($\varphi_{34}$ and $\varphi_{45}$) are related to this unique layer. Any other
correlation function has a partner which includes the same type of junction, for example $(C^{s}_{4,5}$ and
$C^{s}_{2,3})$. Fig.~\ref{2}b shows the enlarged region of Fig.~\ref{2}a in the current interval (0.561,0.564).
It clearly demonstrates that the features of the correlation functions reflect the features of CVC .

\begin{figure}[ht]
 \centering
\includegraphics[height=80mm]{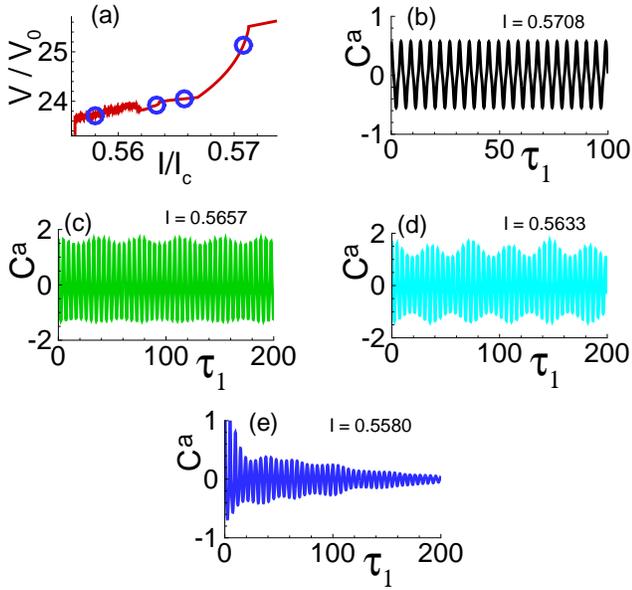}
\caption{(Color online)(a) The CVC of the stack with nine IJJ; (b, c, d, e)  The time dependence of the
autocorrelation function $C^a_5$ in the different parts of the BPR at $I=0.57080, 0.56570, 0.56330$, and
$0.55800$, respectively.}
 \label{5}
\end{figure}
Let us investigate the  charge correlations  in the neighboring layers, using  $C^{c}_{l,l+1} =
<Q_l(\tau)Q_{l+1}(\tau))> =
\lim_{(T_m-T_i)\rightarrow\infty}\frac{1}{(T_m-T_i)}\int_{T_i}^{T_m}{Q_l(\tau)Q_{l+1}(\tau)d \tau}$. Here we
should emphasize that the index $l$ is a layer number. Fig.~\ref{4} presents the dependence  of
$C^{c}_{l,l+1}(I/I_c)$ for different $l$ for the stack with nine IJJ at $\alpha=1$, $\beta=0.2$, with periodic
BC. The negative sign of the $C^{c}_{l,l+1}$ is due to the $\pi$-mode of charge oscillations, so that the
product of positive and negative charges gives us a negative sign for this function. Again, the correlation
functions $C^{c}_{l,l+1}$ come in pairs: there are four pairs $(C^{c}_{4,5},C^{c}_{3,4}),
(C^{c}_{5,6},C^{c}_{2,3}), (C^{c}_{6,7},C^{c}_{1,2}), (C^{c}_{7,8},C^{c}_{9,1})$; here, the correlation function
that stands by itself is $C^{c}_{8,9}$ for layers 8 and 9, which are the farthest from the sp-layer (layer 4).
The correlation functions $(C^{c}_{4,5}$ and $C^{c}_{3,4})$  between $T$ and $Y$ are close to zero, because the
charge on the the sp-layer is close to zero in this interval. The inset shows the enlarged region around point
$B_c$, where the charge on the sp-layer demonstrates its specific dynamics (see Fig.~\ref{3}). The remarkable
fact is that all pairs of current correlation functions $C^{s}_{j,j+1}$ and charge correlation functions
$C^{c}_{l,l+1}$ form a loop reflecting this specific dynamics around this point! It means that the correlation
functions reflect the correlations of phase dynamics amongst all layers and all junctions in the stack, even if
the layers or junctions are far from each other. This is a demonstration of LPW in the system in other language.
An interesting feature is observed in the chaotic region: At transition to the chaotic behavior (point $C_1$),
the values of all correlation functions approach each other. The chaotic region will be discussed in detail
elsewhere.

To distinguish the harmonic oscillations in the BPR from the chaotic behavior, we study the autocorrelation of
the charge on the S-layers by the autocorrelation function $C^a_l = <Q_l(\tau)Q_l(\tau-\tau_1)>=
\lim_{(T_m-T_i)\rightarrow\infty} \frac{1}{(T_m-T_i)} \int_{T_i}^{T_m}{Q_l(\tau)Q_l(\tau-\tau_1)d\tau}$. The
autocorrelation function allows finding the repeating patterns (periodic signals) which have been buried under
noise, or identifying the missing fundamental frequency in a signal, implied by its harmonic content. In
Fig.~\ref{5}a we show the CVC highlighted at $I=0.57080, 0.56570, 0.56330,$ and $0.55800$, where the time
dependence of the autocorrelation function  $C^a_5 = <Q_5(\tau)Q_5(\tau-\tau_1)>$ is investigated. Results are
presented in Figs.~\ref{5}(b)-~\ref{5}(e). The different character of the autocorrelation function reflects the
phase dynamics features in these parts of the BPR. At $I=0.5580$, inside the chaotic domain, we can clearly
distinguish the periodic motion from the chaotic one, in that  $C^a_5$  decays to zero with increasing time
$\tau_1$.

In summary, the phase dynamics of intrinsic Josephson junctions in the high-$T_c$ superconductors is
theoretically studied. We establish a correspondence between the features of current--voltage characteristics,
and the superconducting current in the breakpoint region. We investigated the superconducting current in the
coupled system of Josephson junctions with LPW and clarified the role of the superconducting current
correlations in different junctions, in the formation  of the total CVC. We demonstrated that the correlations
of the superconducting currents in neighboring junctions and the correlations of the charge on superconducting
layers manifest themselves as the features on the CVC,  as a consequence of the phase dynamics in the breakpoint
region. We showed that the correlation analysis is a powerful tool for the investigation of the CVC of the
intrinsic Josephson junctions.

We thank  R. Kleiner, K. Kadowaki, M. Suzuki, I. Kakeya, H. Wang, T. Hatano and F. Mahfouzi  for helpful
discussions. This research was supported by the Russian Foundation for Basic Research, grant 08-02-00520-a. M.
Hamdipour acknowledges financial support from $BLTP$ and $IASBS$.


\begin{thebibliography}{}
\bibitem{ozyuzer}L.Ozyuzer et al, Science {\bf318}, 1291 (2007).
\bibitem{prb}Yu. M. Shukrinov, F. Mahfouzi, N. F. Pedersen, Phys. Rev. B {\bf 75}, 104508 (2007).
\bibitem{prl}Yu. M. Shukrinov, F. Mahfouzi, Phys.Rev.Lett. {\bf 98}, 157001 (2007).
\bibitem{sust1}Yu. M. Shukrinov, F. Mahfouzi, Supercond. Sci.Technol., {\bf 19}, S38-S42 (2007).
\bibitem{kleiner}W. Buckel, R. Kleiner, Superconductivity. Fundamentals and Applications, Wiley-VCH Verlag GmbH $\&$Co, KGaA, (2004).
\bibitem{barone}A. Barone and J. Patterno, Physics and Applications of the Josephson effect, John Wiley and Sons (1982).
\bibitem{likharev}K. K. Likharev, Dynamics of Josephson Junctions and Circuits Gordon and Breach, New York (1986).
\bibitem{irie} A. Irie, Yu. M. Shukrinov, G. Oya, Appl.Phys.Lett. {\bf 93}, 152510, (2008).
\bibitem{prb2}Yu. M. Shukrinov, F. Mahfouzi, M. Suzuki, Phys. Rev. B {\bf 78}, 134521 (2008).
\bibitem{machida99} M. Machida, T. Koyama, and M. Tachiki, Phys. Rev. Lett. 83, 4618 (1999).
\bibitem{machida00} M. Machida, T. Koyama, A. Tanaka and M. Tachiki, Physica {\bf C330}, 85 (2000)
\bibitem{physC2}Yu. M. Shukrinov, F. Mahfouzi, P. Seidel.  Physica {\bf C449}, 62 (2006).

\end{thebibliography}
\end{document}